        \documentstyle[11pt]{article}
\def\ls{\vskip 12.045pt}   
\def\ni{\noindent}        
\def\deg{\ifmmode^\circ \else$^\circ $\fi}    
\def\al{{et\thinspace al.}\ }    
\def\arcs{\ifmmode {'' }\else $'' $\fi}     
\def\arcm{\ifmmode {' }\else $' $\fi}     
\def\buildrel#1\over#2{\mathrel{\mathop{\null#2}\limits^{#1}}}
\def\mper{\ifmmode \buildrel m\over . \else $\buildrel m\over .$\fi}
\def\hper{\ifmmode \rlap.^{h}\else $\rlap{.}^h$\fi}
\def\sper{\ifmmode \rlap.^{s}\else $\rlap{.}^s$\fi}
\def\arcsper{\ifmmode \rlap.{' }\else $\rlap{.}' $\fi}
\def\arcmper{\ifmmode \rlap.{'' }\else $\rlap{.}'' $\fi}
\def\gapprox{$_ >\atop{^\sim}$}     
\def\lapprox{$_ <\atop{^\sim}$}     

\def\ref{\par\noindent\hangindent 20 pt}

\def\mincir{\ \raise -2.truept\hbox{\rlap{\hbox{$\sim$}}\raise5.truept	
\hbox{$<$}\ }}							
\def\magcir{\ \raise -2.truept\hbox{\rlap{\hbox{$\sim$}}\raise5.truept	
\hbox{$>$}\ }}							
\include{BoxedEPS}
\input boxedeps.tex
\SetepsfEPSFSpecial
\HideDisplacementBoxes

\begin {document}

\null
\bigskip\bigskip\bigskip
\bigskip\bigskip\bigskip

\begin{center}

{\bf \LARGE A Determination of the Spectral Index of Galactic Synchrotron 
Emission in the 1-10 GHz range}

\vspace{1.5cm}

{\bf 
P.Platania$^1$,
M.Bensadoun$^2$,
M.Bersanelli$^1$, 
G.De Amici$^3$, 
A.Kogut$^4$, 
S.Levin$^5$,
D.Maino$^6$, 
G.F.Smoot$^2$}\\

\end{center}
\vspace{ 1.5cm}

\ni
$^1$ Istituto di Fisica Cosmica, Consiglio Nazionale 
delle Ricerche, via Bassini 15, 20133 Milano, Italy

\ni 
$^2$ Space Science Laboratory and Lawrence Berkeley 
Laboratory, M/S 50-205, University of California, 
Berkeley CA 94720, USA

\ni 
$^3$ TRW, Space and Technology Division, R1 - 2136, 1 Space Park, Redondo Beach
CA 90278, USA

\ni
$^4$ Hughes STX, Code 685, NASA/GSFC, Greenbelt MD 
20771, USA

\ni 
$^5$ Jet Propulsion Laboratory, Pasadena CA 91109, USA

\ni 
$^6$ SISSA, International School for Advanced Studies,
Strada Costiera 11, 34014, Trieste, Italy

\begin{center}

\vspace{ 3cm}

{\bf DRAFT}
\medskip

( 22--JULY--1997 )

\bigskip

\end{center}

\newpage

\null

\bigskip
\bigskip
\bigskip
\bigskip
\bigskip

{\bf ABSTRACT} ---  We present an analysis of simultaneous multifrequency 
measurements of the Galactic emission in the 1-10 GHz range with 18\deg \, 
angular resolution taken from a high altitude site.
Our data yield a determination of the synchrotron spectral index between 
1.4 GHz and 7.5 GHz of $\alpha_{syn} = 2.81 \pm 0.16$. Combining our data with
the maps from Haslam \al (1982) and Reich \& Reich (1986) we find 
$\alpha_{syn} = 2.76 \pm 0.11$ in the 0.4 - 7.5 GHz range. 
These results are in agreement with the few previously published measurements.
The variation of $\alpha_{syn}$ with frequency based on our results and 
compared with other data found in the literature suggests a steepening
of the synchrotron spectrum towards high frequencies as expected from theory, 
because of the steepening of the parent cosmic ray electron energy spectrum.
Comparison between the Haslam data and the 19 GHz map (Cottingham 1987) also 
indicates a significant spectral index variation on large angular scale.  
Addition quality data are necessary to provide a serious study of these effects.

\section{Introduction}
The detailed study of continuous radio emission allows a direct evaluation
of some important parameters that describe the dynamics and structure of the
Galaxy: the mean intensity of the magnetic field, the spectral index of the 
energy spectrum of cosmic ray electrons and the temperature and density of 
interstellar clouds.    
Additional strong motivation for systematic studies of the diffuse Galactic 
emission arises in conjunction with measurements of the CMB (Cosmic Microwave 
Background): Galactic emission is one of the main sources of unwanted signal 
in CMB observations, and it is unavoidable even in satellite measurements. 
It is then very important to understand in detail the spectral and spatial 
variations of the various components of Galactic emission in order to separate 
them from those due to the CMB. 
For this purpose an experiment dedicated to measuring the low-frequency 
Galactic emission with multifrequency measurements has been carried out in 
1988 from White Mountain, California, as part of a USA-Italy collaboration for 
measuring the spectrum of the CMB in the Rayleigh-Jeans region 
(Smoot \al 1985). 
In this paper we present these previously unpublished data; because of the 
lack of Galactic emission surveys in this frequency range, new results in this 
field are very important.

\subsection{The Galactic emission}
At frequencies lower than $\sim$ 30 GHz Galactic emission is mainly due to 
synchrotron emission from cosmic ray electrons interacting with the Galactic 
magnetic fields and to thermal bremsstrahlung (free-free) emission. 
In the frequency range of our experiment (1-10 GHz) the dominant 
contribution comes from synchrotron radiation (Fig. 1), with free-free 
radiation contributing $\sim 30-50 \%$ on the Galactic plane. 
Our first goal is to determine a mean synchrotron spectral index and 
its possible variation with frequency.
The synchrotron emission arises from relativistic cosmic ray electrons moving 
in the Galactic magnetic fields; for an electron population 
with energy distribution $N(E)$ described by a power law
\begin{equation}
N(E)dE \sim E^{-\delta}dE
\end{equation}
the ensemble synchrotron radiation spectrum in terms of brightness temperature 
is also a power law 
\begin{equation}
T(\nu) \sim \nu^{-\alpha}
\end{equation}
where $\nu$ is the radiation frequency and $\alpha = (\delta+3)/2$.
In the energy range 2 \lapprox $E$ \lapprox 15 GeV (corresponding to 
frequencies between 408 MHz and 10 GHz) the spectral index of the energy 
distribution of the cosmic rays electrons is $\delta \sim$ 3 and then 
$\alpha \sim$ 3. 
For higher electron energies, the energy spectrum steepens and so does the 
radiation spectrum (Banday \& Wolfendale 1991).

\ni The brightness temperature distribution depends on the electron density 
along the line of sight $N(E,l)$ and on the power $P$ emitted by an electron 
of energy $E$ into a magnetic field B
\begin{equation}
T(\nu) \sim \int \int P(\nu, B, E)N(E,l)dEdl 
\end{equation}
Because of the dependence of $B$ and $N(E)$ with the position in the Galaxy, 
one expects $T$ and $\alpha$ to be also functions of the position in the sky.

\section {The Experiment} 
Our analysis is based upon a set of data collected during three nights (6,
8, 10 September 1988), from the White Mountain Barcroft Station, California 
(altitude 3800 m, latitude +37\deg.5; hereafter WM; for a review of the entire
experiment campaign see Smoot \al 1985) with radiometers operating at 1.375, 
1.55, 3.8 and 7.5 GHz (Table 1).
The instruments are total power radiometers, whose output signal, $S$, is 
proportional to the power, $P$, entering the antenna aperture.
The two lower frequencies were covered by a single radiometer 
(Bensadoun \al 1993) which could switch the center frequency of its YIG 
filter. Details on the 3.8 and 7.5 GHz radiometer can be found in De 
Amici \al 1990 and Kogut \al 1990.
Hereafter the signals are expressed in units of antenna temperature 
$T_A = P/kB$, where $P$ is the received power, $k$ is the Boltzmann's constant 
and $B$ is the bandwidth of the radiometer.

During the Galactic scans, each radiometer was tipped 15\deg \, to the East 
and 15\deg \, to the West of the zenith, measuring the signal from the sky 
for 32 seconds in each position.   
We took the difference of the signal in these two positions; in this way
we were able to cancel, at first order, all the isotropic contributions 
(the CMB and the extragalactic sources) and those which give symmetric 
contributions with respect to the zenith angle (the atmospheric and ground 
emissions). (We consider the CMB as isotropic, since the CMB dipole 
contribution in the observed region is negligible, less than 2 mK).   
The differenced signal can thus be written:  
\begin{equation}
T_{A,+15\deg} - T_{A,-15\deg} \simeq \Delta T_{A,Gal} + 
\delta T
\end{equation}
where $\delta T$ includes the second-order contributions from the other terms
and $\Delta T_{A,Gal}$ is the sum of differential ($\pm$ 15\deg) synchrotron 
($\Delta T_{A,syn}$) and free-free ($\Delta T_{A,ff}$) emissions relative to 
the observed sky regions: 
\begin{equation}
\Delta T_{A,gal} = \Delta T_{A,syn} + \Delta T_{A,ff}
\end{equation}

\ni For each night we observed for several hours (see Table 1), thus covering 
significant sections of the sky, with some overlap and redundancy for checking 
systematic effects. 

\ni Calibration measurements were done every hour during 
the experiment using the signal from an ambient temperature load 
(typically $T_{A,amb} \sim 300$ K) and the zenith sky ($T_{A,zenith} \sim 4$ 
K). 
The calibration constant $G$ is 
\begin{equation}
G = \frac{T_{A,amb}-T_{A,zenith}}{S_{amb}-S_{zenith}}
\end{equation}
where $T_{A,amb}$ and $S_{amb}$ are the temperature and the signal when 
the radiometer is looking at the ambient load and $T_{A,zenith}$ and 
$S_{zenith}$ the temperature and the signal when it looks to the zenith sky. 
We calibrated the differential signals between the 15\deg \,E and 15\deg
\,W positions using the value of $G$ resulting from the interpolation between 
two calibration measurements.

\section{Data Analysis}
The integration time is 2 seconds for all the radiometers; we rejected those 
points showing occasional spikes and also those immediately after the change 
in the position of the radiometers. 
The data were then binned in 4\deg \, RA intervals and the 
differences in Eq. (4) were calculated with their statistical errors.

In order to increase information about the synchrotron spectral index, we
compared our data with two sky surveys: the map at 408 MHz of 
Haslam \al (1982) and the map at 1420 MHz of Reich \& Reich (1986).

Sections 3.1 describes data analysis, while Section 3.2 
describes the procedure we used to make the maps comparable with our data. 

\subsection{Differential profiles and corrections}
The WM experiment was designed to have similar antenna beams for all the 
radiometers (see Table 1). For a comparison of data taken at different 
frequencies, we corrected the profiles in order to ``normalize'' the 
different antennas responses to our average antenna beam (Half Power Beam 
Width = 18\deg). 
We generated a ``synthetic'' sky by scaling to our frequencies and convolving 
with the radiometers beam the full-sky 408 MHz Haslam \al map and a catalogue 
of 7400 HII sources at 2.7 GHz (Witebsky 1978) .  
We then calculated for every RA bin $a$ in the profiles the coefficient $\eta$ 
to convert the measured signal into a signal corresponding to an antenna with 
18\deg \, HPBW: 
\begin{equation}
\eta(a) = \frac{T_{\rm 18\deg}(a)}{T_{\rm HPBW}(a)}
\end{equation}
For the points of the differential profile, this translates into the 
conversion coefficient:
\begin{equation}
\eta_{diff} = \frac{\Delta T_{\rm 18\deg}}{\Delta T_{\rm HPBW}}
\end{equation}
where
\begin{equation}
\Delta T_{\rm 18\deg} = T_{\rm 18\deg}(a) - T_{\rm 
18\deg}(b) = T_{\rm 
HPBW}(a)\eta(a) - T_{\rm HPBW}(b)\eta(b) 
\end{equation}
where $a$ and $b$ are generic sky bins whose separation (in RA) is 38\deg \, 
(in fact, observations of points at 15\deg \, from the zenith, carried out at a 
terrestrial latitude of 38\deg \, , correspond to points at 19\deg \, from the 
zenith and declination of 36\deg \, on the celestial sphere).  
Because all the instruments have similar antenna beams, the coefficient 
$\eta_{diff}$ typically varies in the range 0.8-1, thus giving rise to small 
corrections to the data.  

\ni After this correction, the resulting Galactic profiles are directly 
comparable and are shown in Figs. 2a-2d. 
Note the decreasing signal-to-noise ratio with increasing frequency due to 
the spectral behaviour of the synchrotron spectral index (Eq. (2)).

Because the main goal of this work is to evaluate a mean synchrotron 
spectral index in the observed sky region, for all the profiles the 
bremsstrahlung contribution has been evaluated using the catalogue of HII 
sources, convolved with the 18\deg \, ``standard'' antenna beam and scaled at 
the observation frequencies with a spectral index $\alpha_{ff} = 2.1$ 
(Scheffler \& Els\"{a}sser, 1987).
This component has been subtracted leaving the synchrotron component as a 
result (see Section 4.1 for systematic errors arising from the subtraction).

\subsection {The Maps}
The 408 MHz Haslam map is a full-sky survey composed from several data sets at
the same frequency obtained using different telescopes with similar beam size; 
the final angular resolution is 0.85\deg. 
The 1420 MHz Reich \& Reich map covers declinations $\delta >-19\deg$ and the 
angular resolution is 0.6\deg.
From these surveys we extracted the Galactic profiles, corresponding to our
observed sky region, convolved them with a gaussian beam with HPBW = 18\deg, 
and then constructed the differential profiles by simulating the observation 
strategy; the profiles are shown in Figs. 2e and 2f. 
These two surveys have intrinsic uncertainties in zero level and gain 
calibration, the first not affecting our analysis because of the differential 
reduction technique.
The gain calibration error is 10\% for the 408 MHz map and 5 \% for the 
1420 MHz one; these errors have been added in quadrature to the statistical 
errors in both map profiles.

\ni The two maps are in total intensity, while our instruments were 
sensitive only to one linear polarization; to make our data and the maps 
directly comparable we corrected the maps using the linear polarization survey 
of the Galactic background from Brouw \& Spoelstra (1976) convolved with the 
18\deg \, beam. We subtracted from the profiles the component perpendicular to 
the polarization direction of our instruments.                     
The polarization data have a statistical mean error of 0.34 K for the 
0.408 GHz data and 0.06 K for 1.4 GHz. We included these errors as statistical
errors. Finally, the maps, as well as the data, have been corrected for the HII 
contribution and errors arising from this procedure were considered.

\section {Results and Error Analysis}

In order to evaluate the synchrotron spectral index we produced 
temperature-temperature plots, or ``TT-plots''. Such plots display as ordinate 
and abscissa antenna temperatures measured simultaneously at two different 
frequencies on the same region of the sky. 
TT-plots were made for every pair of frequencies between data and 
maps, excluding 1.375-1.55,1.375-1.42, 1.42-1.55 GHz, because they are too 
close to obtain a meaningful result, and 3.8-7.5 GHz for the big errors due to 
the decreased signal-to-noise ratio.
In Fig. 3 all the TT-plots are shown; the resulting synchrotron spectral 
indices are listed in Table 2 with final errors, including the systematic 
effects discussed in Section 4.1. Note the excellent agreement of the results 
obtained using our low frequencies (1.375 and 1.55 GHz) and the 1.42 GHz map, 
which confirms that our data don't have unexpected systematic problems. 
The slope of the best fit to the data, $m$, gives the synchrotron 
spectral index: $\alpha = log(m)/log(\nu_1/\nu_2)$, where $\nu_1$ and $\nu_2$ 
are the frequencies of the two sets of data. 
Table 4 shows the statistical parameters of the TT-plots.
Fig. 4 shows all our results; the spectral indices are evaluated between the 
frequency indicated in the box and the one on the x-axis.
The more significant contribution to the evaluation of $\alpha$ comes from data 
taken in the region close to the Galactic plane, because they have a large 
range of temperature and thus weigh the most in the fit. Thus we can say that
our results are primarily representative of synchrotron emission from the 
Galactic plane.

\subsection{Evaluation of Systematic Effects}

For each of the effects considered in this paragraph, we calculate with the 
TT-plot the spectral index for data with and without a given effect; the 
difference between the two spectral indices is the induced systematic error 
on $\alpha_{syn}$ and we list them in Table 3.
In our analysis, we evaluated all the second order effects producing 
systematic errors; the discussion is divided in two parts, the first one 
dedicated to effects due to the corrections in the Galactic profiles, and 
the second one devoted to the evaluation of instrumental effects.

\ls
We considered the possibility that our HII catalogue was not complete and we 
evaluated the effect of this incompleteness on the results. For this purpose we 
corrected our data with an earlier catalogue of only 900 HII sources and 
evaluated $\alpha_{syn}$: the typical value of the resulting error is 
$\sigma_{alpha,ff} \sim$ 2\%, as shown in Table 3.
We took into account also the effect produced by an error on the free-free 
spectral index, considering values $\alpha_{ff}$ = 2.10 $\pm$ 0.05, obtaining 
typically a percentage error of $\sigma_{\alpha,ind,ff} \sim$ 1\%.
We also evaluated all the indices without the subtraction of 
the HII component: the typical difference between these indices and the 
corrected ones is 0.3. We estimate that in the worse case the HII 
compilation would have 20 \% error resulting in an error of $\sim$ 0.06 in the 
spectral indices which would not change the results significantly.

\ls
Instrumental effects can arise from uncertainties on the beam pattern 
configuration, errors in the pointing direction, variations of 
atmospheric emission and from the possible change of the instrument response 
when tipped between pointing directions.  
It is important to point out that, for the differential reduction technique we 
used, an effect that is constant in time or in sky position would 
not affect the estimate of spectral index; in fact, adding a constant value to 
the data set of an instrument would not change the slope of the fit that is 
used to calculate $\alpha_{syn}$.
An example of constant differential signal is the ground contribution, which 
has not been considered in the analysis.

The uncertainty in the Half Power Beam Width, that has been measured 
during the experiment, is $\pm$ 2\deg. To evaluate the error on 
$\alpha_{syn}$ we used the synthetic sky at each frequency convolved with 
the nominal beams of the experiment (HPBW = $16\deg \, , 
17\deg \, , 18\deg \, , 21\deg \,$) and with the same beams but with 
$\pm$ 2\deg \, in the HPBW. The resulting changes in $\alpha_{syn}$ are the 
errors listed in Table 3 ($\sigma_{\alpha,beam}$ \lapprox 1\%).

The measured uncertainty in the pointing directions is \lapprox 
30\arcm. We made the TT-plot with the Haslam differential profile 
resulting from two simple profiles at distance 29\deg \, in RA 
(instead of 30\deg \,).
The difference between the values of $\alpha_{syn}$ with these two different 
displacements in RA was calculated for every TT-plot with 408 MHz data; a 
mean percentage error ($\sigma_{\alpha,poin}$ = 0.1\%) is obtained.

Fig. 5 shows correlations between data taken in different days for the 
frequencies we analysed; the correlation is very good for the radiometer at 
lowest frequency (1.375-1.55 GHz) and less for the 3.8 GHz one.
To evaluate the effect of data non-repeatability, for each pair of frequencies
we made two TT-plot with data registered September 6th and 8th; with the
differences between each pair of $\alpha_{syn}$ we evaluate a mean percentage 
error on the synchrotron spectral index (2\%).
The origin of this effect could be due to the radiometer position dependence 
but this may arise also from variations of atmospheric emission.
While the water vapor contribution is very small at these frequencies, the 
overall atmospheric radiation varies with position and time for 
changes in pressure profiles (Bersanelli \al 1995) in a way compatible with 
the observed variation in our data. We include this effect in 
$\sigma_{\alpha,atm}$.
   
The uncertainty on the calibration coefficient $G$ was also considered;  
the nature of the resulting error is statistical because it is dominated by the 
uncertainty on the registered value of the ambient temperature. In Table 3 
(typical range $\sim$ 2-5\%) this component is included in the statistical 
data errors.  

\section{Discussion}
We used our results to study the behaviour of the synchrotron spectral index 
with frequency and evaluate a mean synchrotron spectral index in the frequency 
region of this experiment (Section 5.1 and 5.2). The frequency dependence is 
consistent with that expected from the steepening of the Galactic cosmic rays 
electron spectrum. 
In addition, we compared the Haslam map with a map at high frequency (19 GHz) 
to point out the consistent spatial variation of $\alpha_{syn}$ (Section 5.3).

\subsection{Frequency Variation}
As discussed in Section 1, synchrotron radiation arises from cosmic ray
electrons; the interstellar energy spectrum does not have a constant slope, but
steepens at an energy of $ \sim $ 10-20 GeV. In fact cosmic ray electrons lose
energy with different mechanisms: at energies \lapprox 10-20 GeV they escape
from the Galaxy, for E \gapprox 10-20 GeV the prevalent mechanism is 
synchrotron radiation.
Because of the two combined effects, the equilibrium spectrum of the electrons 
changes its slope in this energy range (Webber 1983) corresponding to a 
frequency $\sim$ 10 GHz.
Consequently, the synchrotron spectral index $\alpha_{syn}$, that is related 
with the electrons spectral index $\delta$, increases with increasing frequency
(Lawson\ al 1987, Banday \& Wolfendale 1990 \& 1991).
The results of our analysis (Fig. 4) suggest an initial increase of 
$\alpha_{syn}$ corresponding to our higher frequency 7.5 GHz, although with a 
relatively poor signal-to-noise ratio. At lower frequencies the spectrum is 
more flat, as we expect.
In Fig. 6 we compare our results with the few other published results in our 
frequency range and at higher frequencies and the trend seems to be confirmed.

\subsection{Evaluation of a Mean Synchrotron Spectral Index}
In the frequency range of our measurements, it is possible to evaluate a mean 
synchrotron spectral index in the observed sky region, since there is not an
evident steepening until higher frequencies.
A mean $\alpha_{syn}$ between 1.4 and 3.8 GHz has been estimated using the 
results from the three TT-plots between 1.375, 1.550, 1.42 and 3.8 GHz, as the 
most statistically significant; the result is $\alpha_{syn} = 2.76 \pm 0.09$. 
Next, we calculate the spectral index using all the results, by a 
weighted mean of the indices derived with the TT-plots; this is less 
statistically meaningful because of some redundant information coming from 
different TT-plot (for example the TT-plot between 1.375 and 7.5 and the one 
between 1.375 and 3.8 cover partially the same frequency region) but it gives 
good information on the behaviour of $\alpha_{syn}$ in a larger frequency 
range.  
For each index, the error in the weighted mean is the quadratic sum of
statistical plus pointing and beam errors (the last two errors have an 
independent effect on the TT-plots).
On the other hand, polarization, HII and atmosphere errors were considered 
as systematic, since the effects they have on the different evaluations of
$\alpha_{syn}$ are not independent.
The mean $\alpha_{syn}$ using only our data yields $\alpha_{syn} = 2.81 \pm 
0.16$; adding data from the 0.408 and 1.42 GHz maps, we derive $\alpha_{syn} =
2.76 \pm 0.11$.

\subsection{Spatial Variation}
As we said in Section 1.1, the synchrotron spectral index has also a consistent 
spatial variation due to changes in the electron spectrum and density and in 
the magnetic field intensity with the position in the sky (Eq. (3)). 
As an example, Fig. 7 is a plot of data from the Haslam map versus data from a 
preliminary 19 GHz map (Cottingham 1987; Boughn \al 1990); the best fit line to 
data yields an estimate of $\alpha_{syn}$ = 3.01. However there is a large 
range of possible values of $\alpha_{syn}$ compatible with the dispersion of 
the data; much of the dispersion appears to be correlated with large angular 
scale spectral variation.   
We point out this behaviour of $\alpha_{syn}$ in Fig. 8, where we show the 
ratio between the two maps; more precisely, if $T_1(i)$ is the temperature of 
the i-th pixel in the 408 MHz map and $T_2(i)$ the temperature of the same 
pixel at 19 GHz, the ratio map shows 
\begin{equation}
ratio = \frac{T_2(i)}{T_1(i)} (\nu_2/\nu_1)^{\alpha_{syn}}
\end{equation}      
where we fixed $\alpha_{syn}$ to a mean value of 2.8.
Then, if the synchrotron spectral index changes from this mean value, the 
pixel color in the map changes too; an increase in $\alpha_{syn}$ makes 
color black, a decrease leads to red. 
One can recognize the feature of the North Polar Spur, where the synchrotron 
spectral index is steeper than the average (Lawson \al 1987, Reich \& Reich 
1988). A part from this feature, the spectrum is flatter than the average at 
high galactic latitudes. This is in agreement with Reich \& Reich (1988): they 
found a flattening of the spectrum with increasing latitude in the outer 
Galaxy direction and also, some evidence for a similar effect in the inner 
Galaxy direction (even if only the North Hemisphere data are currently 
available). Also Bloemen \al (1993) found an hardening of the gamma-ray 
spectrum with latitude that translates into a flattening in the radio spectrum.
The reason for the hardening of gamma-ray is still unclear since, from a 
simple diffusion model of the cosmic rays (Ginzburg \& Syrovatskii 1964), a 
spectral steepening with latitude is expected. More complex models (e.g. Reich 
\& Reich 1988, Bloemen \al 1989) such as halo model may explain the observed 
effect with a competition of many different mechanisms (spatial diffusion, 
convection, adiabatic deceleration and energy losses), but the situation is 
very complex and need more data and modeling.  
Finally, we calculated the synchrotron spectral index between the two maps in 
the sky region of our experiment $\alpha_{syn} = 3.00 \pm 0.20$, as shown 
in Fig. 6. 
The 19 GHz map is a preliminary map, thus the ratio between 19 and 0.408 GHz 
only gives an indication of the synchrotron spectral index spatial variation in 
that frequency range, because of the residual errors in the preliminary 19 GHz 
data.

\section{Conclusions}
We have analysed data taken as part of an experiment dedicated to the 
measurement of the low-frequency spectrum of the Galactic continuous emission. 
The radiometers, operating at 1.375, 1.55, 3.8 and 7.5 GHz, observed a sky
region at declination 38\deg. 
The differential reduction technique has allowed us to cancel the first order
contributions from all the isotropic and symmetrical signals received by the 
radiometers, leaving the Galactic signal. 
In order to evaluate the synchrotron spectral index $\alpha_{syn}$, we have 
subtracted from the data the HII component derived from an HII sources 
catalogue.
We have used the TT-plots to calculate the synchrotron spectral index and to
evaluate systematic errors (arising from uncertainties in HII contribution, 
beam pattern configuration, pointing direction and atmospheric emission); we 
have compared our data with the Haslam and Reich \& Reich maps to 
yield an estimate of $\alpha_{syn}$ in the frequency range 0.408-7.5 GHz and
primarily in the region of the Galactic plane.
The general behaviour of $\alpha_{syn}$ in this frequency range suggests a 
steepening of the synchrotron radiation spectrum at frequency $\sim$ 10 GHz, 
already visible at our higher frequency 7.5 GHz.
The mean value of the synchrotron spectral index is 2.81 $\pm$ 0.16 in the 
frequency range of this experiment (1.375-7.5 GHz) and 2.76 $\pm$ 0.11 
including the two maps.

\ni As Fig. 6 shows, it is important to acquire new accurate sets of 
data in the frequency range 1-50 GHz to understand the frequency behaviour of 
the synchrotron spectral index.
In particular, high sensitivity maps of extended regions of the Galaxy at 
sub-degree angular resolution will be extremely important in the context of 
the next generation of CMB experiments, such as the two planned space missions 
MAP and PLANCK Surveyor. 

\ls

\ni {\bf Acknowledgments}

\ni We thank L.Danese, E. Gawiser and C. Paizis for useful discussions and 
suggestions; we also thank J. Aymon for his technical support.
This work was funded in part by the Collaborative Research Grant CRG960175 of 
NATO International Scientific Exchange Program and by the Director, Office of 
Energy Research, Office of High Energy and Nuclear Physics, Division of High 
Energy Physics of the U.S. Department of Energy under contract 
DE-AC03-76SF00098.

\vspace{5cm}

\begin{center}
\begin{tabular}{ccccc}
\multicolumn{5}{c}{\large{\bf Table 1}} \\
\multicolumn{5}{c}{\large{\bf  Radiometers and Observations parameters}} \\
& & & & \\
\hline
\hline
& & & & \\
{\it Frequency [GHz]} & {\it Measured Sensitivity} & 
{\it HPBW$^a$} & {\it Date (1988)} & {\it RA [h])} \\
& & & & \\
\hline
& & & & \\
1.375-1.550 & 18 mK $\rm{Hz^{-1/2}}$ & 18\deg \,  - 
16\deg & 6 Sep & 0-4.3 18.3-24 \\
& & & 8 Sep & 0-5 19-24 \\
3.8 & 13 mK $\rm{Hz^{-1/2}}$ & 17\deg & 6 Sep & 0-4.3 18.3-24 \\
& & & 8 Sep & 0-5 20.3-24 \\
& & & 10 Sep & 21-23 \\
7.5 & 44 mK $\rm{Hz^{-1/2}}$ & 21\deg & 6 Sep & 0-2.3 18.3-24 \\
& & & 8 Sep & 0-5 19-21 \\
& & & 10 Sep & 21-23 \\
& & & & \\
\hline 
& & & & \\
\multicolumn{5}{l}{$a$) The HPBW of the gaussian 
beam that approximates the real one} \\
\end{tabular}
\end{center}

\vspace{2cm}

\begin{center}
\begin{tabular}{ccc}
\multicolumn{3}{c}{\large{\bf Table 2}} \\
\multicolumn{3}{c}{\large{\bf Spectral indices and 
errors}} \\
& & \\
\hline
\hline
& & \\
{\it Frequencies [GHz]} &{\it ~~~~$\alpha_{syn}$} &{\it 
$\sigma_{\alpha,tot}$} \\
& & \\
\hline
& & \\
0.408 - 1.375 & ~~~~2.775 & 0.095 \\
0.408 - 1.42 & ~~~~2.776 & 0.119 \\
0.408-1.55 & ~~~~2.691 & 0.093 \\
0.408 - 3.8 & ~~~~2.731 & 0.079 \\
0.408 - 7.5 & ~~~~2.951 & 0.163 \\
1.42 - 3.8 & ~~~~2.681 & 0.152 \\
1.42 - 7.5 & ~~~~3.057 & 0.252 \\ 
1.375 - 3.8 &  ~~~~2.689 & 0.117 \\   
1.375 - 7.5 & ~~~~3.327 & 0.372 \\   
1.550 - 3.8 & ~~~~2.841 & 0.110 \\ 
1.550 - 7.5 & ~~~~3.609 & 0.508 \\ 
& & \\
\hline
& & \\ 
\end{tabular}
\end{center}

\newpage

\begin{center}
\begin{tabular}{cccccccc}
\multicolumn{8}{c}{\large{\bf Table 3}} \\
\multicolumn{8}{c}{\large{\bf Statistical and Systematic Errors [K]}} \\
& & & & & & & \\
\hline
\hline
& & & & & & & \\
{\it Frequencies [GHz]} & {\it $\sigma_{\alpha,stat}$} & {\it 
$\sigma_{\alpha,beam}$} & {\it $\sigma_{\alpha,poin}$} & {\it 
$\sigma_{\alpha,ind.ff}$} &{\it $\sigma_{\alpha,ff}$} 
& {\it $\sigma_{\alpha,atm}$} & {\it $\sigma_{\alpha,pol}$ ~~} \\
& & & & & & & \\
\hline
& & & & & & & \\
0.408 - 1.375 & 0.064 & 0.032   & 0.003	& 0.017	& 0.025	& 0.055	
& 0~~ \\
0.408 - 1.42  & 0.100 & -	& -	& 0.033	& 0.003	& 0.055	
& 0.001~~ \\
0.408 - 1.550 & 0.060 & 0.039	& 0.003	& 0.012	& 0.022	& 0.054	
& 0.001~~ \\
0.408 - 3.8   & 0.041 & 0.025	& 0.003	& 0.007	& 0.029	& 0.055	
& 0~~ \\
0.408 - 7.5   & 0.130 & 0.002	& 0.003	& 0.049	& 0.062	& 0.059	
& 0~~ \\
1.42 - 3.8    & 0.120 & 0.026 & 0.003	& 0.034	& 0.062	& 0.054	
& 0~~ \\
1.42 - 7.5    & 0.274 & 0.006	& 0.003	& 0.068	& 0.123	& 0.064	
& 0~~ \\
1.375 - 3.8   & 0.061 & 0.07  & 0.003	& 0.036	& 0.028 & 0.054	
& - ~~ \\   
1.375 - 7.5   & 0.320 & 0.001	& 0.003	& 0.120	& 0.129	& 0.066	
& - ~~ \\   
1.550 - 3.8   & 0.069 & 0.039	& 0.003	& 0.035	& 0.033	& 0.057	
& - ~~ \\ 
1.550 - 7.5   & 0.440 & 0.004	& 0.004	& 0.158	& 0.185	& 0.072	
& - ~~ \\ 
& & & & & & & \\
\hline
& & & & & & & \\
\end{tabular}
\end{center}

\vspace{2cm}

\begin{center}
\begin{tabular}{cccc}
\multicolumn{4}{c}{\large{\bf Table 4}} \\
\multicolumn{4}{c}{\large{\bf Statistics}} \\
& & & \\
\hline
\hline
& & & \\
{\it Frequencies [GHz]} & {\it $\chi^2_{\nu}$} & {\it 
$r_S$} & {\it $N_{points}$} \\
& & & \\
\hline
& & & \\
0.408-1.375 & 0.322 & 0.916 & 38 \\
0.408-1.55 & 0.404 & 0.874 & 38 \\
0.408-3.8 & 0.620 & 0.911 & 38 \\
0.408-7.5 & 2.574 & 0.272 & 38 \\
0.408-1.42 & 0.031 & 0.972 & 38 \\
1.42-3.8 & 0.396 & 0.856 & 38 \\
1.42-7.5 & 2.549 & 0.252 & 38 \\
1.375-3.8 & 0.956 & 0.789 & 38 \\
1.375-7.5 & 2.736 & 0.212 & 38 \\
1.55-3.8 & 0.917 & 0.757 & 38 \\
1.55-3.8 & 2.787 & 0.190 & 38 \\
& & & \\
\hline  
& & & \\
\end{tabular}
\end{center}

\begin{center}
\begin{tabular}{ccccc}
\multicolumn{5}{c}{\large{\bf Table 5}} \\
\multicolumn{5}{c}{\large{\bf Other evaluations of $\alpha_{syn}$}} \\
& & & & \\
\hline
\hline
& & & & \\
{\it Frequency range [GHz]} & {$\alpha_{syn}$} & {HPBW} & {\it Reference} & 
{\it Notes} \\
\hline
& & & & \\
0.408-2 & 2.65 $\pm$ 0.05 & 14\deg & Bersanelli \al 1995 & -40 $\le \delta 
\le $ -60 \\
& & & & \\
0.408-31.5 & $\ge$ 2.9 & 7\deg & Kogut \al 1996$^a$ & full sky \\
& & & & \\
1.42-5 & 2.9 $\pm$ 0.3 &  2\deg & Bersanelli \al 1996 & \\
& & & & \\
1.42-10.4 & 3.4 & 5\deg & Bersanelli \al 1996 & \\
1.42-14.9 & 3.0 & 5\deg & Bersanelli \al 1996 & \\
& & & & \\
\hline
& & & & \\
\multicolumn{5}{l}{$a$) We independently estimated the spectral index 
between 0.408 and 31 GHz yielding the same result.} \\ 
\end{tabular}
\end{center}

\begin{figure}[t]
\centerline{
	\ForceHeight{18cm}
	\BoxedEPSF{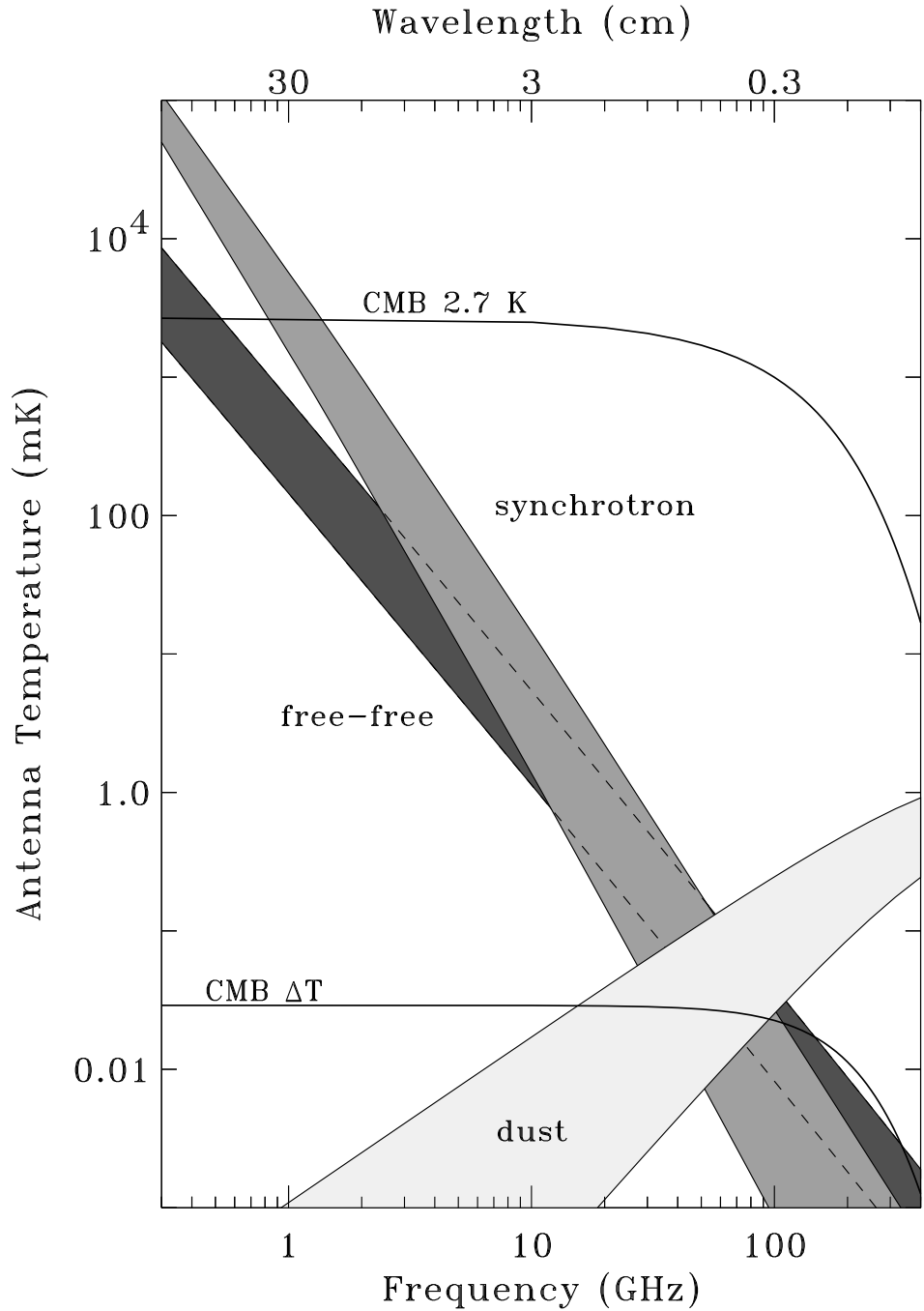}
}
\caption{\em Galactic emission components and CMB spectra for moderate angular 
resolution (7\deg \, HPBW) and galactic latitude $|b| <$  20\deg. The 
shaded regions indicate the range of synchrotron, free-free and dust emission. 
Solid lines indicate the mean CMB spectrum and rms amplitude of anisotropy.}  
\end{figure}

\begin{figure}[t]
\centerline{
	\ForceHeight{18cm}
	\BoxedEPSF{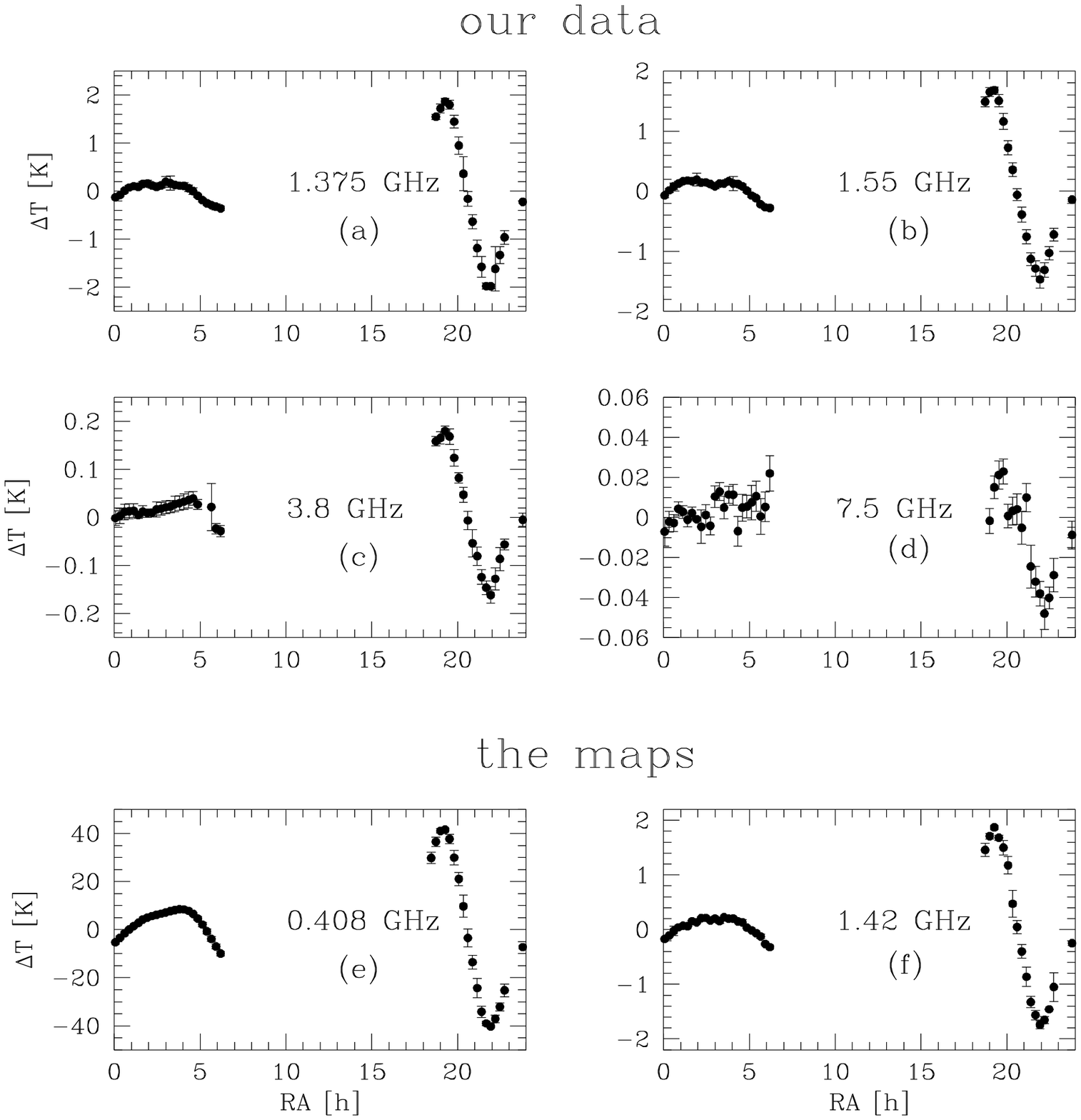}
}
\caption{\em
All profiles. Figs 2a-2d: Differential Galactic profiles derived from our data 
at the frequencies indicated in the plots. Figs 2e-2f: Differential Galactic 
profiles derived from the maps at 408 and 1420 MHz.
}       
\end{figure}

\begin{figure}[h]
\centerline{
	\ForceHeight{18cm}
	\BoxedEPSF{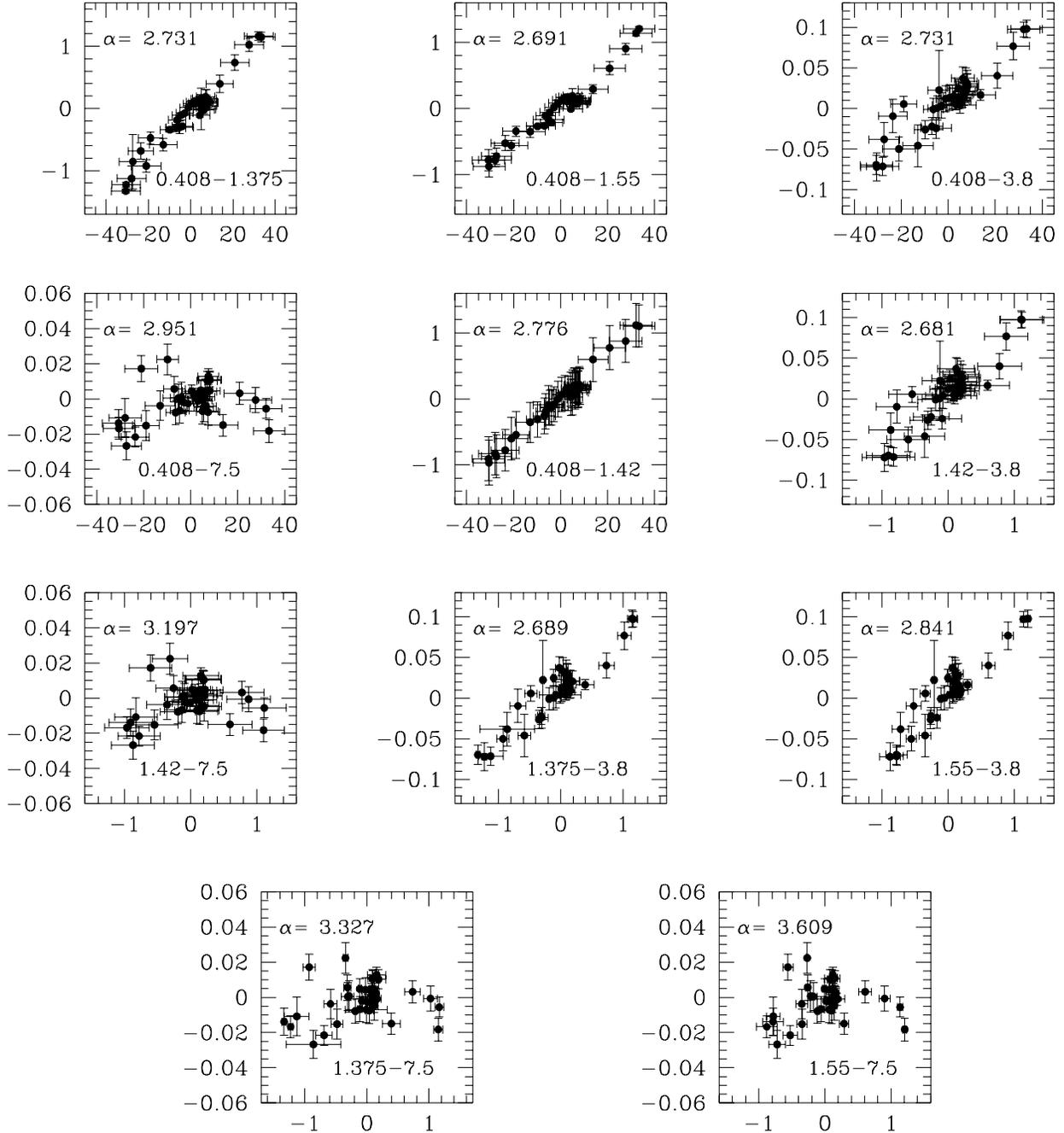}
}
\caption{\em
All the TT-plots are shown. The resulting synchrotron spectral index is 
indicated in every plot, as well as the pair of frequencies (GHz) at which data 
have been taken, the first referring to abscissa and the second to ordinate. 
The units are K on both axis.}
\end{figure}

\begin{figure}[h]
\centerline{
	\ForceHeight{18cm}
	\BoxedEPSF{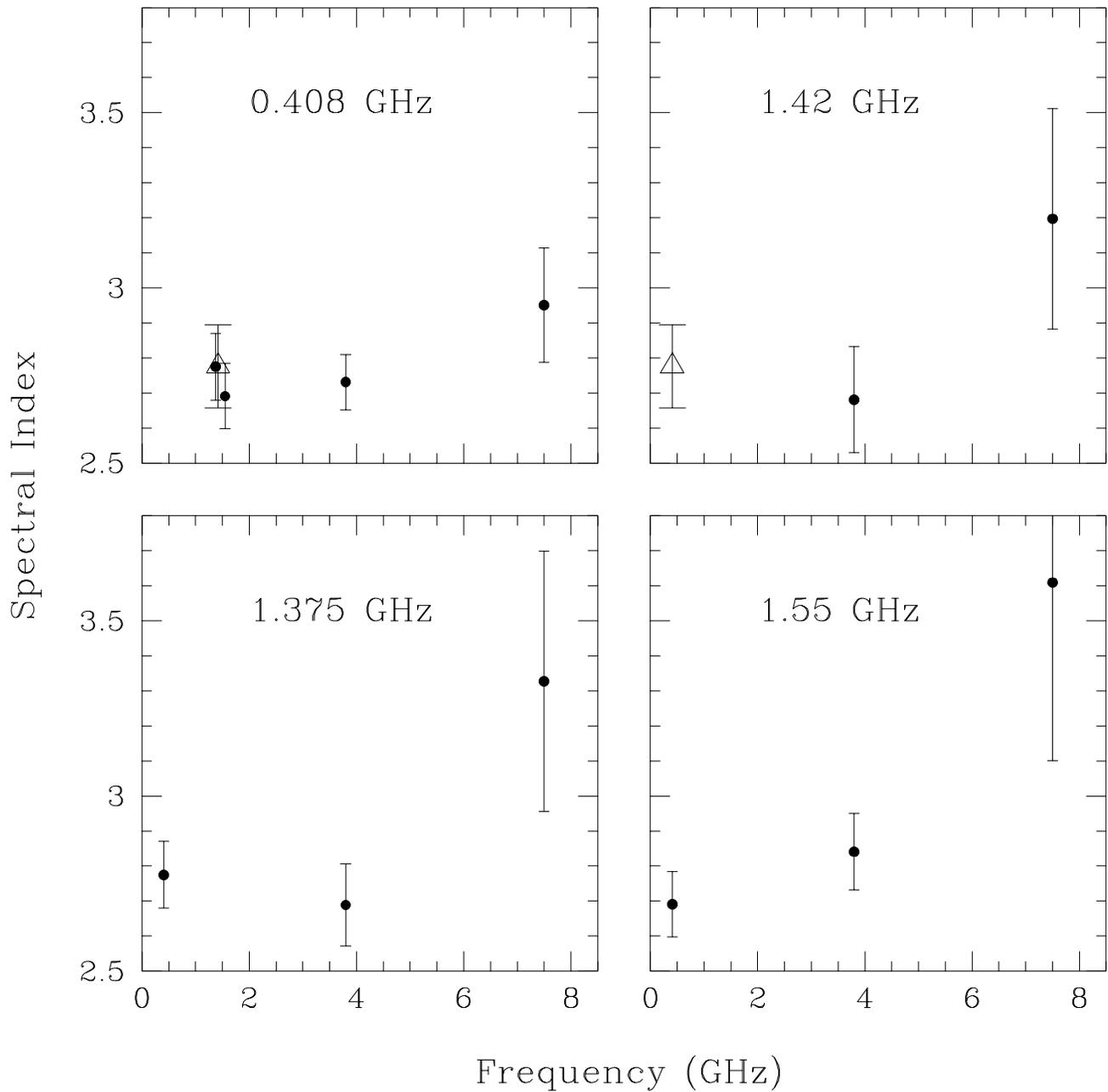}
}
\caption{\em
In each box the spectral indices between the frequency indicated in the box 
and the frequencies on the x-axis are shown. The triangle point refers to our 
evaluation of the synchrotron spectral index between the two maps.}
\end{figure}

\begin{figure}[h]
\centerline{
	\ForceHeight{18cm}
	\BoxedEPSF{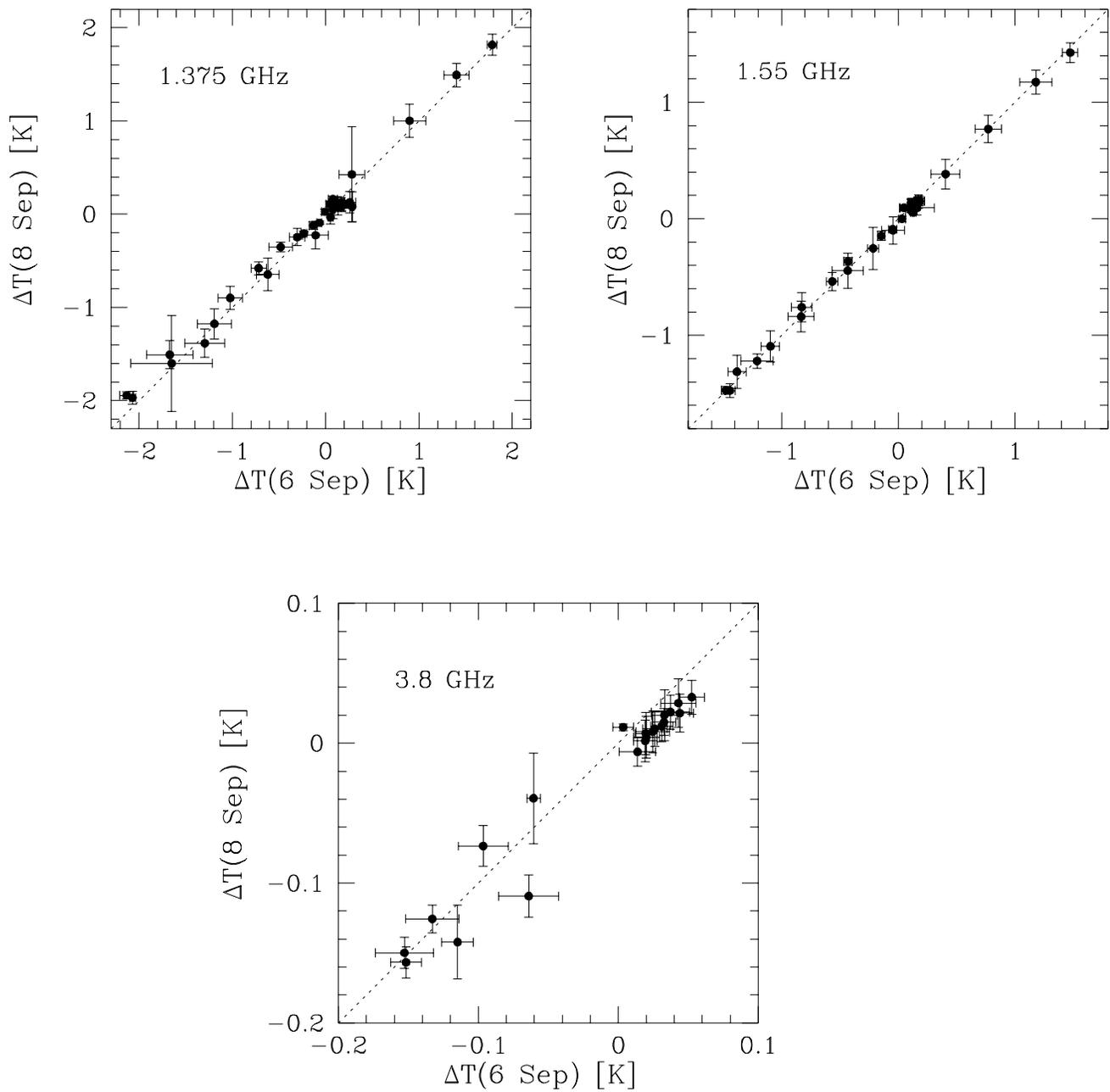}
}
\caption{\em
Correlations between data taken in different days: on the x-axis data taken on 
the 6th of September are plotted, while on the y-axis, those taken on the 8th.}
\end{figure}

\begin{figure}[h]
\centerline{
	\ForceHeight{18cm}
	\BoxedEPSF{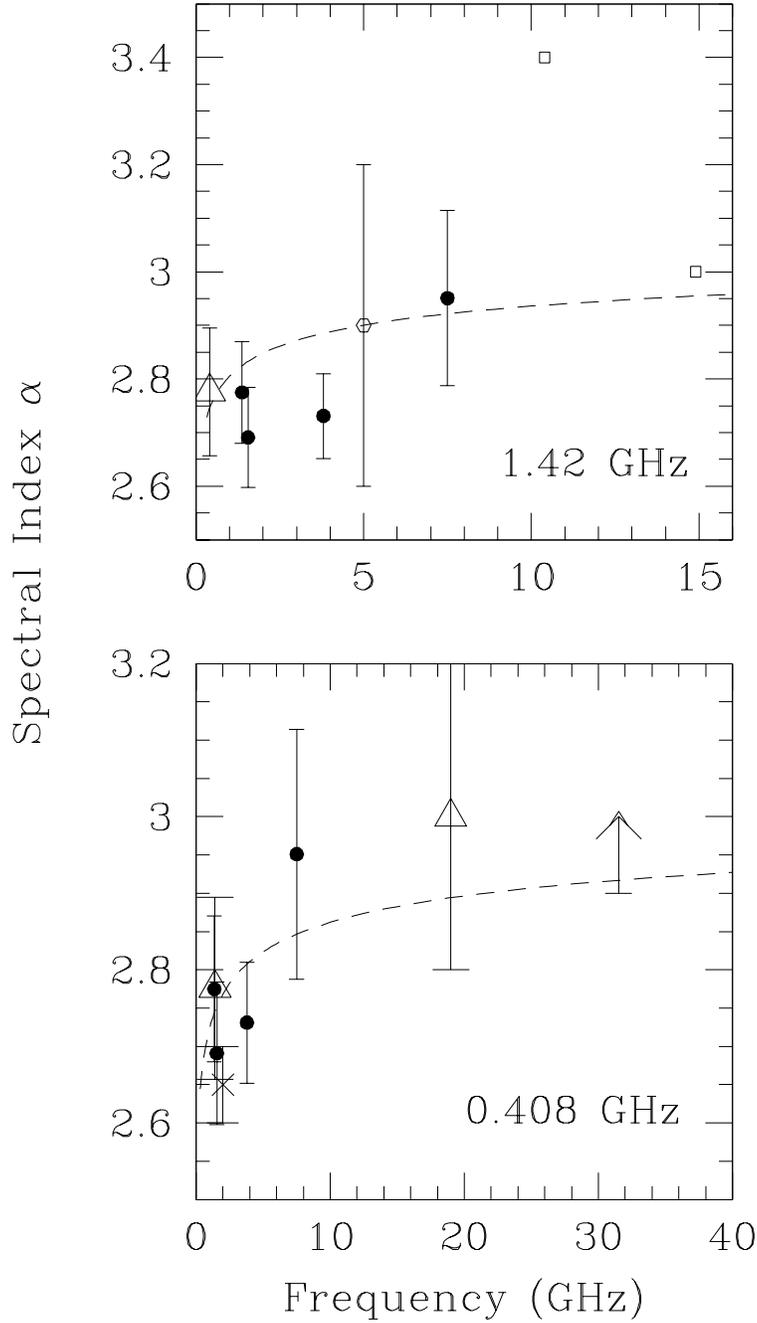}
}
\caption{\em Spectral indices derived in this work are compared with other 
published estimates. The indices are between 1.42 GHz and frequencies on 
the x-axis in the upper plot and between 0.408 GHz and frequencies on
the x-axis in the lower plot. Full circles: this work. Empty triangles: our 
evaluations of $\alpha_{syn}$ from the maps at 0.408, 1.42 and 19 GHz. See 
Table 5 for other points's references. The smoothed dashed lines are naive 
calculation of the spectral index based on the local cosmic ray electron 
spectrum.} 
\end{figure}

\begin{figure}[h]
\centerline{
	\ForceHeight{12cm}
	\ForceWidth{12cm}
	\BoxedEPSF{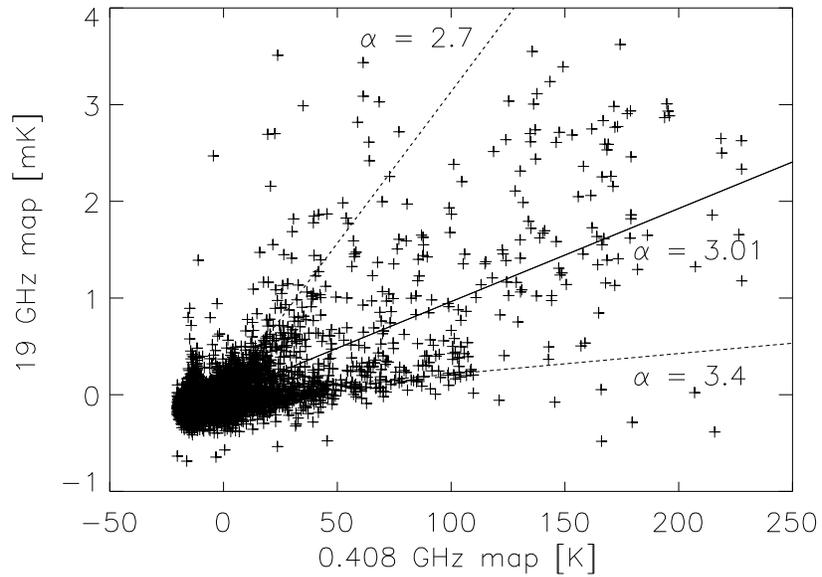}
}
\caption{\em TT-plot with all-sky data from 408 MHz and 19 GHz maps. In the 
plot best fit to data is shown (solid line) and upper and lower limits 
(dashed lines) to possible spectral index values. An offset has been 
subtracted to 19 GHz data; some data have negative values because the mean has 
been subtracted to calculate the best fit.}
\end{figure}

\begin{figure}[t]
\centerline{
 	\ForceHeight{18cm}
	\BoxedEPSF{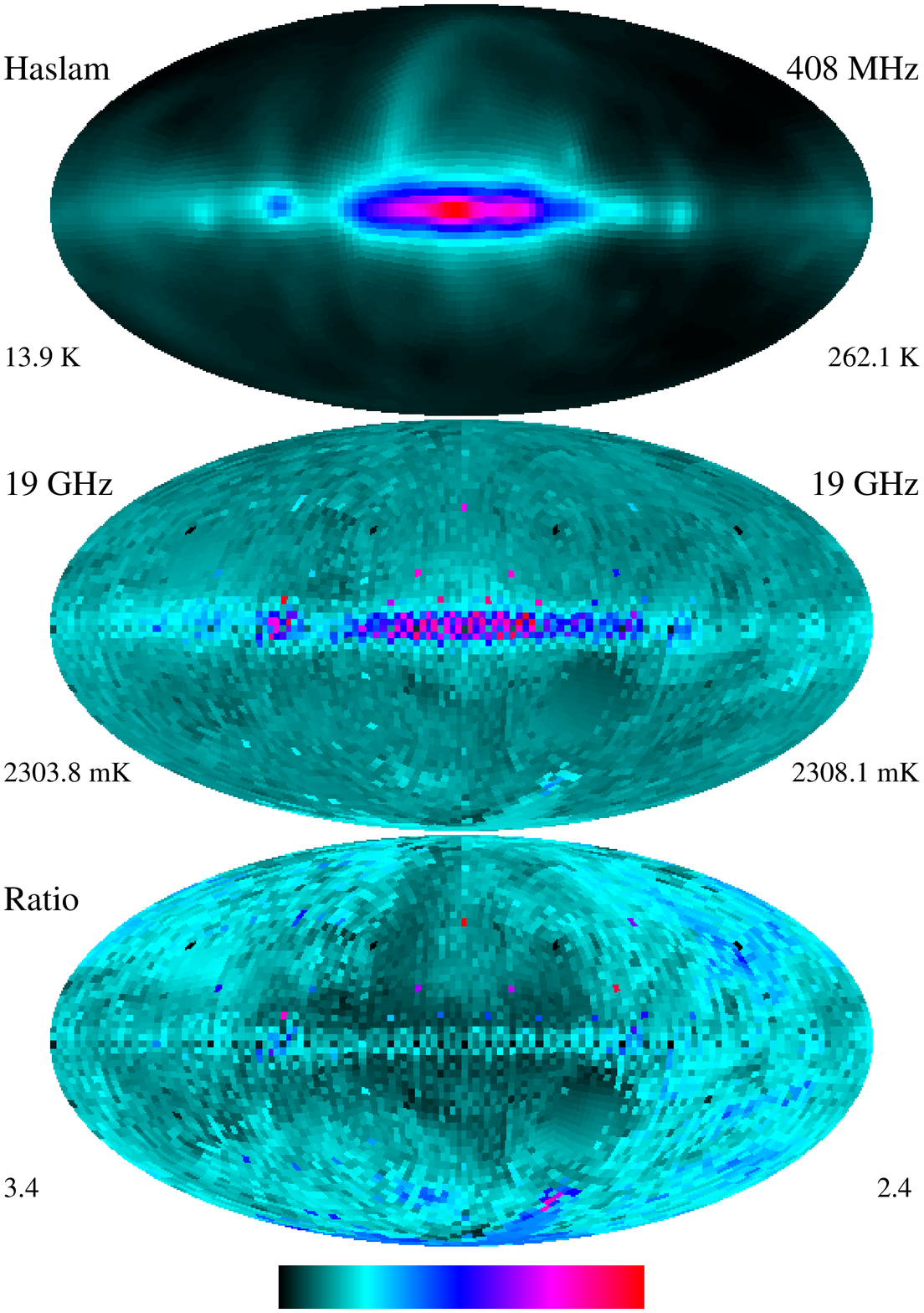}
}
\caption{\em Top: Haslam \al 408 MHz full-sky map. The minimum and maximum 
temperatures are labeled at the bottom of the map. Middle: preliminary 19 GHz 
map (Cottingham 1987, Boughn \al 1990). The two smoothed regions (one in 
the southern hemisphere on the right, the other one in the northern hemisphere 
on the left) are not covered by the survey and the blank pixels have been 
replaced with the average temperature from surrounding regions. The minimum 
and maximum temperature at the bottom include the offset of the survey. 
Bottom: Ratio map between the 0.408 and 19 GHz maps. The bottom labels refer 
to the maximum and minimum spectral index.}    
\end{figure}

\newpage

\ni {\bf References}

\ref Banday, A. J. \& Wolfendale, A. W. 1990, {\em MNRAS}, {\bf 245}, 182.

\ref Banday, A. J. \& Wolfendale, A. W. 1991, {\em MNRAS}, {\bf 248}, 705.

\ref Bensadoun, M. \al 1993, {\em ApJ}, {\bf 409}, 1. 

\ref Bersanelli, M. \al 1995, {\em Astro. Lett. and Communications}, {\bf 32}, 
7. 

\ref Bersanelli, M. \al 1995, {\em ApJ}, {\bf 448}, 8.

\ref Bersanelli, M., Bouchet, F.R., Efstathiou, G., Griffin, M., Lamarre, J.M.,
Mandolesi, N., Nordgaard-Nielsen, H.U., Pace, O., Polny, J., Puget, J.L., 
Tauber, J., Vittorio, N., Volonte', S. 1996, {\em COBRAS-SAMBA: report on the 
phase A study}, ESA. 

\ref Bloemen, H. \al 1993, {\em Astron. Astrophys.}, {\bf 267}, 372. 

\ref Boughn, S. P.\al 1990, {\em Rev. Sci. Instr.}, {\bf 61}, 158.

\ref Brouw, W. N. \& Spoelstra, T. A. Th. 1976, {\em Astron. Astrophys. Suppl.
Ser.}, {\bf 26}, 129. 

\ref Cottingham, D. A. 1987, {\em PhD thesis}, Princeton Univ.

\ref De Amici, G. \al 1990, {\em APJ}, {\bf 359}, 219.

\ref Ginzburg, V. L. \& Syrovatskii, S. I. 1964, {\em The origin of the Cosmic Rays}, 
Oxford, Pergamon Press.

\ref Haslam, C. G. T. \al 1982, {\em Astron. Astrophys. 
Suppl. Ser.}, {\bf 47}, 1.
 
\ref Kogut, A. \al 1990, {\em ApJ}, {\bf 355}, 102.

\ref Kogut, A. \al 1996, {\em ApJ}, {\bf 460}, 1.

\ref Lawson, K. D. \al 1987, {\em MNRAS}, {\bf 225}, 307. 

\ref Reich, P. \& Reich, W. 1986, {\em Astron. Astrophys. Suppl. Ser.}, {\bf 
63}, 205. 

\ref Reich, P. \& Reich, W. 1988, {\em Astron. Astrophys. Suppl. Ser.}, {\bf 
74}, 7.

\ref Scheffler, H. \& Els\"{a}sser, H. 1987, {\em Physics of the Galaxy and 
Interstellar Matter}, Springer-Verlag, Heidelberg.

\ref Smoot, G. F. \al 1985, {\em ApJ Lett.}, {\bf 291}, L23.

\ref Webber, W. R. 1983, in {\em Composition and origin 
of cosmic rays;
Proceedings of the Advanced Study Institute}, Erice, 
Italy, June 20-30, 1982,
Dordrecth, D. Reidel Publishing Co., 83.

\ref Witebsky, C. 1978, not published.

\end{document}